\documentclass{LMCS}

\def\doi{8 (1:20) 2012}
\lmcsheading%
{\doi}
{1--23}
{}
{}
{Feb.~\phantom01, 2011}
{Mar.~\phantom08, 2012}
{}

\usepackage{enumerate,hyperref}
\usepackage{amsmath}
\usepackage{amssymb}
\usepackage{proof}
\usepackage{goedelsat}

\theoremstyle{plain}
\theoremstyle{plain}

\begin{document}

\title[Theorem proving for prenex $\Gd$]%
{Theorem proving for prenex G\"odel logic with~$\det$: \\
checking validity and unsatisfiability}
\thanks{{\lsuper{a,b,c}}Partly supported by FWF grant~P22416, WWTF grant~WWTF016,
FWF START Y544-N23, and
Eurocores-ESF/FWF grant 1143-G15 (LogICCC-LoMoReVI)}

\author[M.~Baaz]{Matthias Baaz\rsuper a}
\address{{\lsuper a}Department of Discrete Mathematics and Geometry, TU Vienna}
\email{baaz@logic.at}

\author[A.~Ciabattoni]{Agata Ciabattoni\rsuper b}
\address{{\lsuper{b,c}}Department of Computer Languages, TU Vienna}
\email{\{agata,chrisf\}@logic.at}
%\thanks{Supported by WWTF grant~WWTF016.}

\author[C.~G.~Ferm\"uller]{Christian G.\ Ferm\"uller\rsuper c}
\address{\vskip-6 pt}
%\email{chrisf@logic.at}
%\thanks{Supported by Eurocores-ESF/FWF grant 1143-G15 (LogICCC-LoMoReVI)}


%******************************************************
\begin{abstract} First-order
G\"odel logic with the projection operator $\det$ ($\Gd$) is an important
many-valued as well as intermediate logic. In contrast to classical logic,
the validity and the satisfiability problems of $\Gd$ are not directly
dual to each other.
We nevertheless provide a uniform, computational treatment of both problems 
for prenex formulas
by describing appropriate translations into sets of order clauses that can be
subjected to chaining resolution.
For validity a version of Herbrand's Theorem allows us to show the soundness
of standard Skolemization. For satisfiability the translation involves  
a novel, extended Skolemization method.  
\end{abstract}

\maketitle

\section{Introduction}

In classical logic efficient, resolution based theorem proving is a two step process.
To prove the validity of an arbitrary first-order formula~$F$ 
we first translate its negation into  a Skolemized, purely
universal conjunctive normal form of $\neg F$. This normal form directly
corresponds to ``logic free''  syntax, namely to a set of clauses~$cl(\neg F)$
that can be subjected to Robinson's celebrated resolution mechanism to test for unsatisfiability.
The well attested efficiency of resolution, compared to other proof search methods,
is due to the combination of the unification principle (the existence of most general unifiers) and simple atomic cuts as the
only inference steps. The original formula $F$ is valid if and only if
the empty clause, representing contradiction, is derivable from~$cl(\neg F)$ 
in this manner.

Note that in classical logic testing the validity of $F$
is equivalent to testing the unsatisfiability of~$\neg F$. 
This duality is lost in 
the logic~$\Gd$ that we will consider here, 
namely G\"odel logic enriched by the projection operator~$\det$~\cite{baaz96}.
%the extension of G\"odel logic that we consider here. This is $\Gd$, ~i.e.,
%G\"odel logic enriched by the projection operator $\det$.
There are indeed $\Gd$-formulas~$F$ that are unsatisfiable, 
in the sense that there is
no interpretation in $\Gd$ that assigns the designated truth value~$1$ to $F$, 
and nevertheless $\neg F$ is not valid in $\Gd$. 
The importance of G\"odel logic is emphasized by the fact that 
it naturally turns up in a number of different contexts; among them  
fuzzy logic~\cite{hajek}, intermediate logics \cite{TT84}, the provability logic of 
Heyting arithmetic \cite{visser} and logic programming \cite{LPV}. 
In these contexts, both the validity problem and the satisfiability problem are of
interest. In particular the latter is often crucial for applications, 
for instance to detect inconsistencies in the knowledge base of fuzzy rule-based systems~\cite{CR10}.
%On the other hand, both, the validity problem and the
%satisfiability problem are of great potential interest for applications. The latter, for instance,
%is crucial to detect inconsistencies in the knowledge base of fuzzy rule-based systems \cite{CR10}.
The operator $\det$, which maps  $\det G$ to the designated
truth value~$1$ if the value of $G$ equals~$1$, and to $0$ otherwise,
greatly increases the expressive power of G\"odel logic and its applicability.
However,
it renders the resulting logic more complicated. For instance,
checking whether a formula of $\Gd$ is valid or satisfiable
is undecidable already in the prenex and monadic case,
i.e.\ when considering formulas with only unary predicates and 
no function symbols where a string of quantifiers precedes a
quantifier free part; in contrast, without~$\det$
satisfiability for prenex formulas is decidable, see~\cite{BCP09,BCF01}.

Our aim is to provide a uniform treatment of the validity and 
the satisfiability problem  for first-order $\Gd$ in as close analogy to 
classical logic as possible.

In contrast to propositional G\"odel logic with or without $\det$ (e.g. \cite{Fiorino,LW07}), efficient 
automated theorem proving at the first-order level seems to be beyond the
current state of the art, if possible at all. Thus it is reasonable
to consider appropriate non-trivial fragments.
Here we focus on the {\em prenex} fragment of $\Gd$.
We describe an efficient translation of such formulas into 
sets of order clauses that can then be subjected to chaining
resolution~\cite{BGjacm,BGcade} to test for unsatisfiability.
A central challenge here is to prove the soundness of
(appropriate versions of) Skolemization.
 
The results below mainly bring together and round off 
what we have presented in preliminary form in two conference 
papers, \cite{BCF01} and~\cite{BF10}.
The current paper is organized as follows. Section~\ref{sec:Gd} provides
formal definitions and basic facts about~$\Gd$. 
Section~\ref{sec:skolem} is devoted 
to Skolemization for prenex~$\Gd$. 
This requires a separate and different treatment
of the validity and the satisfiability case. In Subsection~\ref{ss:skolemval}
we prove that the standard Skolemization
method that replaces strong quantifier occurrences by newly introduced
Skolem terms preserves validity for prenex formulas in~$\Gd$. The
proof involves a version of Herbrand's theorem as well as a rather
general statement about ``reverse Skolemization'',
which is of independent interest. For testing (un-)satisfiability
we cannot proceed as in classical logic, but rather introduce a novel 
extended form of Skolemization that, in addition to
replacing weak quantifier occurrences by Skolem terms, introduces
a fresh monadic predicate symbol. 
In Section~\ref{sec:TP} we demonstrate that the results of Section~\ref{sec:skolem}
enable a translation of prenex $\Gd$-formulas into corresponding clause forms
of a particular kind, namely order clauses that are to be interpreted
in the theory of dense linear orders with endpoints.
To achieve a reasonably efficient translation process we
use definitional normal forms 
that introduce new predicate symbols for non-atomic subformulas.
We show that both the validity
and the satisfiability problem for prenex~$\Gd$ can be reduced to
proving (un-)satisfiability of corresponding sets of order clauses.
For the final step of theorem proving we rely on known results about 
so-called ordered chaining resolution~\cite{BGcade,BGjacm}. To render
the paper self-contained we will explicitly state in
Subsection~\ref{ss:resolution} which inference
rules and corresponding soundness and completeness result are needed
in our specific case.

%*********************************************************
\section{G\"odel Logic with $\det$} \label{sec:Gd}

First-order G\"odel logic~$\G$, sometimes also called intuitionistic
fuzzy logic \cite{TT84} or Dummett's~{\LC} (eg.\ in \cite{avr91,Gab72}, 
referring to \cite{dummett}),
arises from intuitionistic logic by
adding the axiom of linearity $(P \Impl Q) \ \Or \ (Q \Impl P)$
and the axiom
$\A x (P(x) \Or Q^{(x)}) \Impl (\A x P(x)) \Or Q^{(x)}$
($\Or$-shift), where the notation $A^{(x)}$ indicates that there is no free
occurrence of~$x$ in~$A$.

Semantically G\"odel logic can be viewed as an infinite-valued
logic where the real interval $[0,1]$ is taken as the set of truth values.%
\footnote{For more 
information about G\"odel logic---its winding history, importance, 
variants, alternative semantics and proof systems---see, e.g.,
\cite{BPZ07,BCF03,hajek,P10,HBMFL_goedel}.}
An {\em interpretation}~$\I$ consists of a non-empty domain~$D$
and a valuation~$v_{\I}$ that maps constant symbols and object
variables to elements of $D$ and $n$-ary function symbols to functions
from $D^n$ into $D$; $v_{\I}$ extends in the usual way to a function
mapping all terms of the language to an element of the domain. Moreover,
every $n$-ary predicate symbol~$p$ is mapped to a function $v_{\I}(p)$ 
of type $D^n \mapsto [0,1]$. The truth-value of an atomic formula ({\em atom})
$p(t_1,\ldots,t_n)$ is defined as
\[
\val{\I}{p(t_1,\ldots,t_n)} = v_\I(p)(v_\I(t_1),\ldots,v_\I(t_n)).
\]
For the truth constants
$\bot$ and $\top$ we have $\val{\I}{\bot} = 0$ and $\val{\I}{\top} = 1$.

The semantics of propositional connectives is given by\\[-1ex]
\[
%\small
\begin{array}{lcl}
\val{\I}{A \And B}  = \min(\val{\I}{A}, \val{\I}{B}), & \ \ \ \ &
         \val{\I}{A \Or B}  = \max(\val{\I}{A}, \val{\I}{B}), 
\end{array}
\]
\[
\begin{array}{c}
\val{\I}{A \Impl B}  =
 \begin{cases}
   1            & \mbox{if } \val{\I}{A} \leq \val{\I}{B}   \\
   \val{\I}{B}       & \mbox{otherwise}.
  \end{cases} 
\end{array}
\]
Henceforth we will consider the following abbreviations:
$\neg A$ for $A \Impl \f$ and
$A \Iff B$ for $(A \Impl B) \And (B \Impl A)$. Therefore
\[
\begin{array}{lc@{}l}
\val{\I}{\neg A}  =
 \begin{cases}
   1            & \msp \mbox{if } \val{\I}{A} = 0\\
   0            & \msp \mbox{otherwise}
\end{cases}
& \   &
 \val{\I}{A \Iff B} =
\begin{cases}
1                   &  \msp \mbox{if } \val{\I}{A}\,{=}\,\val{\I}{B}\\
\min(\val{\I}{A},\val{\I}{B}) & \msp  \mbox{otherwise}. 
\end{cases}
\end{array}
\]
For quantification we define the {\em distribution}
of a formula $A$ with respect to a free variable $x$
in an interpretation $\I$ as
$\distI(A(x)) = \{\val{\I[d/x]}{A(x)} \mid d \in D\}$, where
$\I[d/x]$ denotes the interpretation that is exactly as $\I$,
except for insisting on $v_{\I[d/x]}(x)=d$. Similarly we will
use $\I[\tup{d}/\tup{x}]$ for the interpretation arising from~$\I$
by assigning the domain element $d_i$ in $\tup{d} = d_1, \ldots, d_n$
to the variable $x_i$  in $\tup{x} = x_1, \ldots, x_n$ ($1 \le i \le n$). 
The universal and existential quantifiers  correspond to the
infimum and supremum, respectively,
in the following sense:
\[
\val{\I}{\forall x A(x)} = \inf \distI(A(x))
\ \ \ \ \ \
\val{\I}{\exists x A(x)} = \sup \distI(A(x)).
\]

\subsection{The projection operator $\det$}
%%%%%%%%%%%%%%%%%%%%%%%%%%%%%%%%%%%%%%%%%%

Following~\cite{baaz96}, we enrich the language of $\G$ by adding
the unary operator $\det$ with the following meaning:
\[
\val{\I}{\det A}  =
 \begin{cases}
   1            & \mbox{if } \val{I}{A} = 1    \\
   0            & \mbox{otherwise}.
 \end{cases}
\]
The resulting logic---denoted as $\Gd$---is strictly more expressive 
than $\G$. $\det$ allows to recover classical reasoning inside
``fuzzy reasoning''  in  a very simple and natural manner: if all
atoms are prefixed by $\det$ then $\Gd$ coincides with
classical logic. However, the expressive power of $\det$
goes considerably beyond this. In particular, observe that in general $\det \E x P(x)
\Impl \E x\det{}P(x)$ is not valid in $\Gd$. (There are interpretations
$\I$ such that $\val{\I}{\E x P(x)} = 1$ although $\val{\I[d/x]}{P(x)} < 1$
for all domain elements~$d$.).

\begin{defi}
A formula $A$ is \emph{valid} in $\Gd$ (in symbols: $\modelsg A$) if $\val{\I}{A}=1$ for all
interpretations~$\I$. $A$ is \emph{(1-)satisfiable}
 (in symbols: $A \in \oSAT$)
if $\val{\I}{A}=1$ for at least one interpretation~$\I$. Every such interpretation 
is called a \emph{model} of~$A$.
\end{defi}

\begin{rem} In $\G$ and $\Gd$ validity as well as 1-satisfiability of a formula
depend only on the \emph{relative order} of the truth values of 
atomic formulas, but not on their specific values.
\end{rem}

\begin{rem} 
%As already pointed out in the introduction, 
%with a concrete example: 
In contrast to classical logic, 
in $\Gd$ validity and 1-(un)satisfiability are not dual 
concepts.%
\footnote{The duality is preserved when considering the notion of 
\emph{positive satisfiability}: a formula $A$ is positively satisfiable
if there exists an interpretation $\I$ such that
$\val{\I}{A}>0$. However this notion is less natural than 1-satisfiability.
In particular a formula can be positively satisfiable without admitting a model.} 
For instance, the formula  $\neg(B \And \neg\det B)$ is not valid in~$\Gd$,
although $B \And \neg\det B$ is unsatisfiable. 
%For instance, the (propositional) formula $$(C \prec D) \wedge (D \prec C) \wedge \neg \det C$$
%where $A \prec B$ abbreviates $((B\Impl A) \Impl B)$, is not 1-satisfiable
%but its negation is not valid (notice that for each interpretation
%$\I$, $\I(A \prec B) = 1$ if and only if $\I(A) < \I(B)$ or $\I(A) = \I(B) = 1$ and 
%$\I(A \prec B) =  \I(B)$ otherwise).
\end{rem}

Since $\Gd$ does not contain the identity predicate, the following
version of the L\"owenheim-Skolem theorem is easily
obtained, just like for classical or intuitionistic logic.
\begin{prop}[\cite{BPZ07}] \label{prop:LS}
Every 1-satisfiable formula of $\Gd$
has a model with countably infinite domain.
\end{prop}

We list a few valid schemes of~$\Gd$ that will be used in later sections.
Recall that the notation
$A^{(x)}$ indicates that there is no free occurrence of $x$ in~$A$.

\begin{lem}
\label{lemma:properties}
Let $A$ and $B$ be formulas of $\Gd$. 
\begin{enumerate}[\em(1)]
\item $\modelsg \det A \Impl A$
\item $\modelsg \det (A \Or B) \Iff (\det A \Or \det B)$
\item $\modelsg A(t)^{(x)} \Impl \E x A(x)$
\item $\modelsg \det \A x A \Iff \A x \det A$
\item $\modelsg \A x (A^{(x)} \Impl B(x)) \Impl (A^{(x)} \Impl \A x B(x))$ 
\item $\modelsg \A x(B(x) \Impl A^{(x)})) \Impl (\E x B(x) \Impl A^{(x)})$ 
\end{enumerate}
\end{lem}

%***********************************************************
\subsection{Prenex Fragment}
%***********************************************************

The prenex fragment of a logic is the set of all closed formulas of
the form $\Q_1 x_1 \ldots \Q_n x_n P$, where $P$ is quantifier free and
$\Q_i \in \{\A,\E\}$ for  $1\le i \le n$.
Like in intuitionistic logic, also in G\"odel logic (with or without $\det$)
quantifiers cannot be shifted arbitrarily. Indeed the following classical
quantifiers shifting laws do not hold:
\begin{iteMize}{$\bullet$}
\item $(\A xA \Impl B^{(x)}) \Impl \E x(A \Impl B^{(x)})$ 
\item $(B^{(x)} \Impl \E x A) \Impl \E x(B^{(x)} \Impl A)$.
\end{iteMize}
As a consequence arbitrary formulas are not equivalent to prenex formulas, in general.
Nevertheless the prenex fragment of~$\Gd$ is quite expressive.
By encoding the classical theory of two equivalence relations one can show
that both the validity and the 1-satisfiability problem are undecidable, 
see \cite{BCP09,BCF01}. 
In fact the prenex fragment of $\Gd$ is undecidable already
in the monadic case; i.e., when considering only unary predicates 
and no function symbols. This should be contrasted with the decidability
of satisfiability of prenex monadic~$\G$~\cite{BCP09} and
of the validity problem for 
the prenex fragment of intuitionistic logic~\cite{DV}.

%***********************************************
%*********************************************************
\section{Skolemization for prenex $\Gd$} \label{sec:skolem}
%**************************************************
%********************************************

Loosely speaking, Skolemization with respect to validity is the replacement 
of strong quantifiers in a formula by fresh function symbols.
Here positive occurrences of universal quantifiers
and negative occurrences of existential quantifiers are called \emph{strong}; 
the other quantifier occurrences are called \emph{weak}.
Skolemization with respect to satisfiability replaces the weak quantifiers in
a formula instead.
It is, of course, always possible to instantiate bound 
variables and delete corresponding quantifier occurrences.
But the aim is to obtain a formula that is valid or 1-satisfiable
if and only if the original formula is valid or 1-satisfiable, respectively.
In classical logic this is achieved by 
replacing each variable occurrence~$y$ that is bound by a strong
(weak) quantifier by a Skolem term~$f(x_1,\ldots,x_n)$
if $y$ is in the scope of the weak (strong) quantifier occurrences
$\Q_1 x_1$, \ldots,  $\Q_n x_n$. We refer to this transformation as 
``standard Skolemization''.
We emphasize the fact that standard Skolemization is not sound for intuitionistic logic,
not even for its prenex fragment (see, e.g., \cite{Mint}).

Below we describe how and why Skolemization 
with respect to validity as well as to satisfiability can be achieved
for the prenex fragment of~$\Gd$. More precisely, 
we show that the standard Skolemization method 
of classical logic also works for prenex formulas of~$\Gd$ 
in the context of validity, but fails with respect to
1-satisfiability. For the latter we present a novel Skolemization method
where, in addition to Skolem terms that replace weak quantifier
occurrences, a fresh monadic predicate symbol is introduced. 
%  MOVE TO INTRODUCTION:
% The results below are an extension of those in\cite{BCF01,BF10}.

%***********************************************
\subsection{Validity} \label{ss:skolemval}
%***********************************************

As in the case of intuitionistic logic, standard Skolemization does not preserve validity in $\G$ (and therefore in $\Gd$).
For instance the formula $$\forall x \neg \neg  A(x) \Impl \neg \neg \forall x A(x) $$ is 
not valid in $\G$%
\footnote{Any interpretation with domain $\{d_1,d_2,\ldots\}$ and
$\inf_{i\ge 1}\val{\I[d_i/x]}{A(x)} =0$ but $\val{\I[d_i/x]}{A(x)} > 0$, for all $d_i$,
 is not a model for the formula.} 
while its Skolemized version 
$\forall x \neg \neg  A(x)  \Impl \neg \neg  A(c)$ is valid.  
In this section we show that the standard Skolemization method 
is nevertheless sound for the prenex fragment of $\Gd$ with respect to validity.

\begin{defi}
\label{def:Skolem}
Let $F=\Q_1 y_1 \dots \Q_n y_n P$, with $\Q_i \in \{\A, \E\}$
be a (prenex) formula, where $P$ is quantifier free.
The operator $\Phi(\cdot)$, to be applied from outside to inside, is 
defined as follows:
\begin{iteMize}{$\bullet$}
\item $\Phi(\E x A(x))$ = $\E x\Phi(A(x))$;
\item $\Phi(\A x A(x))$ = $\Phi(A(f(\tup{y})))$,
     where $f$ is a fresh (Skolem) function symbol and $\tup{y}$ are the
      free variables in $\A x A(x)$; if there are no such variables then
     $\Phi(\A x A(x))$ = $\Phi(A(c))$, for a fresh (Skolem) constant symbol~$c$;
\item $\Phi(A)$ = $A$, if $A$ is quantifier free.
\end{iteMize}
The {\em Skolem form} of $F$, denoted by
$\E\tup{x} P^S(\tup{x})$%
\footnote{The notation hides the fact that
the Skolem form also depends on the quantifier prefix. However, below, the
context will always provide the relevant information.}%
, is $\Phi(F)$.
\end{defi}

Our aim is to provide a constructive proof that 
$$\modelsg \Q_1 y_1 \dots \Q_n y_n P \quad \IFF \quad \modelsg \E\tup{x} P^S(\tup{x}).$$
The first step towards establishing the (more difficult) left-to-right direction of this equivalence
is to prove Herbrand's theorem for prenex $\Gd$ (see \cite{BCF01} or \cite{BPZ07}). 

\begin{defi}
Let $P$ be a formula. The {\em Herbrand universe} $U(P)$
of  $P$ is the set of all variable free terms 
that can be constructed from the set of function symbols and constants
occurring in  $P$. To prevent $U(P)$ from being finite or empty we add a constant
and a function symbol of positive arity if no such symbols appear
in~$P$.

The {\em Herbrand base} ${\B}(P)$ is the set of atoms
constructed from the predicate symbols in $P$ and
the terms of the Herbrand universe. A {\em Herbrand expansion} of $P$
is a disjunction of instances of $P$ where free variables
are replaced with terms in~$U(P)$.
\end{defi}

The following lemma relies on the fact that
the truth value of any formula $P$ of $\Gd$ under a given interpretation
only depends on the ordering of the respective values of
atoms occurring in~$P$.

\begin{lem}
\label{lemma:prop}
      Let $P$ be a quantifier free formula of $\Gd$. For every
      interpretation~$\I$ such that $\val\I{P} < 1$ and every 
      real number $c$, where $0 < c < 1$, 
      there is an interpretation $\Ic$ such that $\val\Ic{P} \le c$.
\end{lem}
\proof
Let $p_1, \dots , p_k$ the atomic formulas in $P$ that do not evaluate to
$0$ or $1$ under $\I$. Without loss of generality, assume 
that $\val\I{p_1} \eors_1 \dots \eors_k \val\I{p_k}$
where each $\eors_i$ is either $=$ or $<$. $\Ic$ is defined by 
assigning (possibly) new values $\val\Ic{p_i}$ to $p_i$ such that 
$\val\Ic{p_1} \eors_1 \dots \eors_k \val\Ic{p_k}$ and 
$\val\Ic{p_k} \le c$ and keeping
the values $0$ or $1$ for the remaining atoms.
The claim is easily proved by structural induction on~$P$.
\qed

\begin{thm} \label{univH}
Let $P$ be any quantifier-free formula of $\Gd$. If
$\modelsg \exists \tup{x} P(\tup{x})$ then there exist
tuples $\tup{t_1}, \dots \overline{t_n}$
of terms in $U(P)$,
such that
$\modelsg \bigvee_{i=1}^{n} P(\tup{t_i})$.
\end{thm}
\proof
Let $A_1, A_2, \dots $ be a non-repetitive
enumeration of (the infinite set) $\B(P)$.
We construct a ``semantic tree'' $\T$; i.e., a
systematic representation of all possible order types of interpretations of $A_i$.
$\T$ is a rooted tree whose nodes appear at levels.
Each node at level~$\ell$ is labelled with  an expression, called
{\em constraint}, of the form 
$$\lab{\ell}{\pi} \defeq 0 \eors_0 A_{\pi(1)} \eors_1 \dots 
               \eors_{\ell-1} A_{\pi(\ell)} \eors_{\ell} 1,$$
where $\eors$ is either $=$ or $<$ and
$\pi$ is a permutation of $\{1, \ldots, \ell\}$.
We say that an interpretation~$\I$ of $P(\tup{x})$
{\em fulfills} the constraint $\lab{\ell}{\pi}$ if\\[-2.4ex]
$$0 \eors_0 \val\I{A_{\pi(1)}} \eors_1 \dots 
               \eors_{\ell-1} \val\I{A_{\pi(\ell)}} \eors_{\ell} 1$$
holds. We say that the constraint
$\lab{\ell+1}{\pi'} \defeq 0 \eors_0 A_{\pi'(1)} \eors_1 \dots 
               \eors_{\ell} A_{\pi'(\ell+1)} \eors_{\ell+1} 1$
{\em extends} $\lab{\ell}{\pi}$ if every interpretation fulfilling
$\lab{\ell+1}{\pi'}$ also fulfills $\lab{\ell}{\pi}$.

$\T$ is constructed inductively as follows:
\begin{iteMize}{$\bullet$}
  \item The root of~$\T$ is at level~$0$ and is labelled with the constraint $0<1$.
  \item Let $\nu$ be a node at level $\ell$ with label $\lab{\ell}{\pi}$.
  \begin{iteMize}{($\ast_1$)}
   \item[($\ast_1$)] Given $\lab{\ell}{\pi}$, if there is 
         an instance $P(\tup{t})$ of $P(\tup{x})$, such that 
        for every interpretation $\I$ that fulfills~$\lab{\ell}{\pi}$ 
        we have
        $\val\I{P(\tup{t})}=1$, 
        where the  atoms of $P(\tup{t})$ are among $A_1, \ldots, A_{\ell}$, 
        then $\nu$  is a leaf node of~$\T$.
   \end{iteMize}
        Otherwise, for each constraint $\lab{\ell+1}{\pi'}$
        that extends $\lab{\ell}{\pi}$
        a successor node $\nu'$ labelled with this constraint
        is appended to $\nu$  (at level $\ell+1$).
\end{iteMize}
By the definition of $\T$ the following holds:
  \begin{iteMize}{($\ast_2$)}
   \item[($\ast_2$)] For every interpretation $\I$ of $\B(P)$ there is a
       branch of $\T$ such that $\I$ fulfills all constraints 
      at all nodes of this branch.
   \end{iteMize}
Two cases are to be considered:
\begin{desCription}
   \item\noindent{\hskip-12 pt\bf $\T$ is finite}:\ Let $\nu_1, \ldots, \nu_m$ be the leaf nodes of~$\T$.
      Then by ($\ast_1$) and ($\ast_2$) we obtain 
        $\modelsg \OR_{i=1}^{m} P(\tup{t_i})$,
      where $P(\tup{t_i})$ is an instance of $P(\tup{x})$ such that
      $\val\I{P(\tup{t_i})}=1$ for all interpretations $\I$ that fulfill
      the constraint at~$\nu_i$.
   \item\noindent{\hskip-12 pt\bf $\T$ is infinite}:\  By K\"onig's lemma, $\T$ has an infinite branch.
      This implies that there is an interpretation $\I$ such that
      $\val\I{P(\tup{t_i})} < 1$ for every tuple $\tup{t_i}$ of terms
      of $U(P)$. By Lemma \ref{lemma:prop} 
      there is an interpretation $\Ip$ satisfying all constraints in the branch,
      such that  $\val\Ip{P(\tup{t_i})} < c < 1$. Hence
      $\val\Ip{\E \tup{x}P(\tup{x})} < 1$, which
      contradicts the assumption that $\modelsg \E \tup{x} P(\tup{x})$. \qed
\end{desCription}

\noindent The following lemma states sufficient conditions for a logic to
admit {\em de-Skolem\-ization} (also known as reverse Skolemization).
By this we mean the re-introduction
%The following lemma establishes sufficient conditions for a logic to
%allow {\em reverse Skolemization} (see \cite{BCF01} or \cite{BPZ07}). By this we mean the re-introduction
of quantifiers in Herbrand expansions. These conditions are fulfilled by $\Gd$ and indeed, by most of the
fuzzy logics in the sense of \cite{hajek}.
Here, by a {\em logic}~$\lo$ we mean a set of formulas 
that is closed under modus ponens.
We call a formula $P$ {\em valid} in $\lo$ (and write: $\modelsl P$)
if $P \in \lo$.

%%%%%%%%%%%%%%%%%%%%%%%%%%%%%%%%%%%%%%%%%%%%%%%%%%%%%
%%%%%%%%%%%%%%%%% DE-SKOLEMIZATION %%%%%%%%%%%%%%
%%%%%%%%%%%%%%%%%%%%%%%%%%%%%%%%%%%%%%%%%%%%%%%%%%%%
\begin{lem}
\label{reskolemization}
Let $\lo$ be a logic satisfying the following properties:\\[-3.4ex]
\protect
\begin{enumerate}[\em(1)]
\item $\modelsl P(t)^{(x)} \IMPLIES \models P(x)$ 
\item $\modelsl Q \Or P  \IMPLIES 
        \modelsl P \Or Q$
\item $\modelsl (Q \Or P) \Or R \IMPLIES 
         \modelsl Q \Or (P \Or R)$ 
\item $\modelsl Q \Or P \Or P \IMPLIES 
       \modelsl Q \Or P$ 
\item $\modelsl P(y) \IMPLIES \modelsl \A x [P(x)]^{(y)}$
\item $\modelsl P(t) \IMPLIES \modelsl \E x P(x) $
\item $\modelsl  \A x (P(x) \Or Q^{(x)}) \IMPLIES \modelsl (\A x P(x)) \Or Q^{(x)}$
\item $\modelsl  \E x (P(x) \Or Q^{(x)}) \IMPLIES \modelsl (\E x P(x)) \Or \E x Q^{(x)}$.
\end{enumerate}
Let $\E \tup{x} P^S(\tup{x})$ be the Skolem form of
$\Q_1 z_1 \dots \Q_n z_n P(z_1, \dots ,z_n)$, where $\Q_i \in\{\forall,\exists\}$.
For all tuples of terms
$\tup{t_1}, \ldots, \tup{t_m}$
of the Herbrand universe of $ P^S(\tup{x})$\\[-2ex]
$$\modelsl \OR_{i=1}^{m} P^S(\tup{t_i}) \ \  \IMPLIES \ \ 
\modelsl \Q_1 z_1 \dots \Q_n z_n P(z_1, \ldots ,z_n).$$
\end{lem}

\Proof\hspace{-0.5ex}%
\footnote{Our de-Skolemization procedure follows the proof of 
the Second Epsilon Theorem by Hilbert and Bernays~\cite{HB}. 
A more detailed modern 
presentation of de-Skolemization for classical logic using
Gentzen's sequent calculus
can be found in~\cite{BHW}. 
Our task is to show that conditions 1--8 are sufficient 
to guarantee the soundness of the procedure and
to correct an error in the proof of this lemma as 
it appeared in~\cite{BCF01,BPZ07}.
}. 
Let $\SKT$ be the set of all instances in $\OR_{i=1}^{m} P^S(\tup{t_i})$
of Skolem terms in $\E \tup{x} P^S(\tup{x})$. 
% Note that each $s \in \SKT$
% is connected to a unique quantifier occurrence $\A z_i$
% in $\Q_1 z_1 \dots \Q_n z_n P(z_1, \dots ,z_n)$ 
% via $\E \tup{x} P^S(\tup{x})$.
We define the following order on $\SKT$:  $s \vord t$ iff  
either $s$ is a proper subterm of $t$ 
or $s=f(t_1,\ldots,t_a)$ and $t=g(t'_1, \ldots,t'_b)$, where
the Skolem term $f(x_1,\ldots,x_a)$ in $\E \tup{x} P^S(\tup{x})$
replaces a variable 
$z_i$ in $\Q_1 z_1 \dots \Q_n z_n P(z_1, \dots ,z_n)$ and
$g(t'_1, \ldots,t'_b)$ replaces a variable $z_j$ 
in $\Q_1 z_1 \dots \Q_n z_n P(z_1, \dots ,z_n)$ such that $i < j$.
(Skolem constants are treated as $0$-ary Skolem functions here.)
 As we will show below, this order 
guarantees that one can re-introduce universal quantifiers 
in $\OR_{i=1}^{m} P^S(\tup{t_i})$ at the
appropriate positions by replacing \emph{maximal} terms at each
corresponding step. 

Starting with $\OR_{i=1}^{m} P^S(\tup{t_i})$ and
working from right to left with respect to the original
quantifier prefix $\Q_1 z_1 \dots \Q_n z_n$, we stepwise (re-)introduce 
quantifier occurrences at individual disjuncts using the algorithm below.
We will use $\pp{i}$ to denote a disjunct in the formula obtained
at some stage of the procedure: the quantifier prefix $\overline{\Q^i}$ is
either empty or else is $\Q_k z_k \dots \Q_n z_n$ for some $1\le k \le n$.
Referring to $\pp{i}$ we will use $\Q^i_+ z$ to denote the quantifier occurrence
immediately preceding $\overline{\Q^i}$ in $\Q_1 z_1 \dots \Q_n z_n$.
If $\overline{\Q^i}$ is empty then $\Q^i_+ z=\Q_n z_n$. (If $\overline{\Q^i}$
is already the full prefix $\Q_1 z_1 \dots \Q_n z_n$ 
then $\Q^i_+$ remains undefined.)
Note that  $\Q^i_+ z$ denotes the quantifier 
occurrence that is to be introduced next at $\pp{i}$. 

\begin{desCription}
\item\noindent{\hskip-12 pt\bf Step 1}:\ The following is repeated as long as possible:\\
    Pick a disjunct $\pp{i}$ in the current formula where 
    $\Q^i_+ z = \exists z_k$ for some $1\le k \le n$. 
     Replace  $\pp{i}(t)$ by $\exists z_k \pp{i}(z_k)$, 
    where $t$ is the term that occurs at those positions in $\pp{i}$ where 
    $z_k$ occurs in $P(z_1, \ldots ,z_n)$. 
\item\noindent{\hskip-12 pt\bf Step 2}:\ Remove all redundant copies of identical disjuncts, if any.
\item\noindent{\hskip-12 pt\bf Step 3}:\ Let $t$ be a maximal term (with respect to $\vord$) in $\SKT$
      and let
     $\pp{i}(t)$ be some disjunct of the current formula where~$t$ occurs.    
     $\Q^i_+ z = \forall z_k$ for some $1\le k \le n$. 
     Replace  $\pp{i}(t)$ by $\forall z_k \pp{i}(z_k)$. Goto step 1.
\end{desCription}

\noindent We claim that the above algorithm converges at 
$\Q_1 z_1 \dots \Q_n z_n P(z_1, \dots ,z_n)$ and 
that the validity of the current formula is preserved at
each stage. To see why conditions 1--8 guarantee that this is the case
we refer to the three steps separately. 

\begin{desCription}
\item\noindent{\hskip-12 pt\bf Ad Step 1}:\ Let $\OR_{i=1}^{\ell} \pp{i}$ be the current formula.
      If  there is no disjunct $\pp{i}$  where 
    $\Q^i_+ z = \exists z_k$ for some $1\le k \le n$ then step 1 is empty.
     Otherwise observe that 
  $$\modelsl \OR_{j=1}^{j=i-1} \pp{j} \Or  \pp{i}(t)  \Or 
                  \OR_{j=i+1}^{j=\ell} \pp{j} 
  $$   
  implies
  $$\modelsl \E z_k (\OR_{j=1}^{j=i-1} \pp{j} \Or \pp{i}(z_k)  \Or 
                  \OR_{j=i+1}^{j=\ell} \pp{j})  
  $$   
  by assumption 6. Note that throughout the procedure the current formula 
  remains closed,   i.e.\ without free occurrences of variables.
  Therefore $z_k$ only occurs free in $\pp{i}(z_k)$ and
  we can apply assumption 8, combined with 2 and 3, to obtain
  $$\modelsl \OR_{j=1}^{j=i-1} \pp{j} \Or \E z_k\pp{i}(z_k)  \Or 
                  \OR_{j=i+1}^{j=\ell} \pp{j} 
  $$   
  as required.

\item\noindent{\hskip-12 pt\bf Ad Step 2}:\ Validity is preserved by assumptions 2, 3, and 4. 
  Moreover, since the original quantifier occurrences $\Q_k z_k$
  are re-introduced at their correct positions, the algorithm terminates
  with the original formula $\Q_1 z_1 \dots \Q_n z_n P(z_1, \ldots ,z_n)$,
  if steps 1 and 3 are sound.

\item\noindent{\hskip-12 pt\bf Ad Step 3}:\ We claim that (i) every maximal term $t$  occurs only
  in a single disjunct $\pp{i}(t)$ of the current formula. Moreover,
  (ii) $\Q^i_+ z= \forall z_k$ for some $1\le k \le n$ and 
   $t$ only occurs at positions where $z_k$ occurs in the original formula. 
  Note that if these claims are true then
  $$\modelsl \OR_{j=1}^{j=i-1} \pp{j} \Or  \pp{i}(t)  \Or 
                  \OR_{j=i+1}^{j=\ell} \pp{j} 
  $$   
  implies
  $$\modelsl \A z_k (\OR_{j=1}^{j=i-1} \pp{j} \Or \pp{i}(z_k)  \Or 
                  \OR_{j=i+1}^{j=\ell} \pp{j})  
  $$   
  by assumption 1 and 5. This allows us to apply assumption 7, combined with 2 and 3, 
  to obtain
  $$\modelsl \OR_{j=1}^{j=i-1} \pp{j} \Or \A z_k\pp{i}(z_k)  \Or 
                  \OR_{j=i+1}^{j=\ell} \pp{j}
  $$   
  as required.

It remains to prove claims~(i) and~(ii).
Since $t = f(t_1,\ldots,t_n)$ is an instance of a Skolem term
$f(x_1,\ldots,x_m)$, $t$ is connected
to the universal variable~$z_j$ in the original formula that has been
replaced by $f(x_1,\ldots,x_m)$ in $\E \tup{x} P^S(\tup{x})$.
Let $\pp{j}(t)$ be a disjunct of the current formula in
which $t$ occurs. 
Because of step~1 we know that $\Q^j_+ z = \forall z_k$ for some
$1 \le k \le n$. Any position $p$ at which $t$
occurs in $\pp{j}(t)$ is such that $z_k$ occurs in the original formula 
$\Q_1 z_1 \dots \Q_n z_n P(z_1, \ldots ,z_n)$ at~$p$ for the following
reasons.
First, by maximality $t$ cannot be a subterm of any term
that replaces one of $z_1, \ldots, z_n$ in the current formula.
Moreover, by the definition of $\vord$, $t$ cannot replace
any variable $z_j$ where $j<k$.
This settles claim~(ii).

Regarding claim (i), 
suppose that there were two
disjuncts $\pp{i}$ and $\pp{j}$ in the current formula in
which $t$ occurs. The maximality of  $t=f(t_1,\ldots,t_n)$, claim (ii),
and the fact that we have already re-introduced all 
quantifier occurrences to the right of 
$\forall z_k$ in  $\Q_1 z_1 \dots \Q_n z_n$
implies that the variables in the Skolem term 
$f(x_1,\ldots,x_m)$ correspond to
precisely those variables in $\E \tup{x} P^S(\tup{x})$ that are
still instantiated by terms from the Herbrand universe in
the current formula. But this means that at each position corresponding
to an occurrence of $x_i$ ($1\le i \le m$) in $\E \tup{x} P^S(\tup{x})$ we
find the same term $t_i$ in the current formula.
Therefore $\pp{i}$ and $\pp{j}$ must
be identical and thus have been contracted into a single disjunct
at the preceding step~2.\eop
\end{desCription}

\begin{cor}
\label{cor:SKOval}
Let $F = {\AND_{1 \le i \le m}}A_i$ where $A_i$ are prenex formulas of $\Gd$.
Then
$$
  \modelsg F 
\quad \IFF \quad \modelsg {\AND_{1 \le i \le m}} \E \tup{x} A_i^S(\tup{x}).$$
\end{cor}

\proof
$(\Rightarrow)$ Follows from the laws of quantification that hold in~$\Gd$.\\
$(\Leftarrow)$ Using Theorem~\ref{univH} and Lemma~\ref{reskolemization}
we argue as follows:
\[
\begin{array}{rl@{\quad}l}
  & \modelsg \E \tup{x} A_i^S(\tup{x}) & \\[1.2ex]
\IMPLIES&  \modelsg \OR_{i=1}^{n} A_i^S(\tup{t_i}) \mbox{ for appropriate }
               \tup{t_1}, \ldots, \tup{t_m}
 &  \mbox{ by Theorem \ref{univH}} \\[1.2ex]
\IMPLIES& \modelsg A_i 
  &  \mbox{ by Lemma \ref{reskolemization}} 
\end{array}
\] 
Since $\modelsg A$ iff for each
$i = 1, \dots, m$, $\modelsg A_i$, this concludes the proof. 
\qed

%****************************************************************
\subsection{Satisfiability}  \label{ss:skolemsat}
%****************************************************************

In contrast to the validity case, standard Skolemization for
satisfiability (where weakly quantified variables are replaced by
Skolem terms) does \emph{not} preserve 1-satisfiability, even for prenex formulas
of $\Gd$. This is due to the fact that to be 1-satisfiable, an
existentially quantified formula does not need to evaluate to~$1$ for 
any of its instances. Rather, it is sufficient that the supremum of the
distribution, i.e., of 
the truth values taken by the instances, is~1. For instance, 
the following formula is 1-satisfiable:
$$\E x p(x) \And \A y \neg \det p(y).$$
A model with domain $\{d_i \mid i \ge 1\}$ is obtained 
by setting $v_\I(p)(d_i) = 1 - \frac{1}{i}$ for all~$i\ge 1$. 
On the other hand the standard Skolemized form
for this formula, $p(c) \And \A y \neg \det p(y)$, is not 1-satisfiable.
(The example can easily be made prenex by moving the universal quantifier
to the front.)
%However, we will slightly widen our focus to \emph{conjunctions} of prenex formulas, anyway.

Below we show that a Skolem form with respect to satisfiability
for ({conjunctions} of) prenex formulas of $\Gd$ can nevertheless be achieved 
by introducing an additional monadic predicate symbol. 
The resulting formulas will have no existential quantifier and will be 1-satisfiable if and only
if the original formulas are. Given a prenex formula its Skolem form will be defined in two steps:
\begin{enumerate}[(1)]
\item We will first introduce a suitable formula, whose {\em only} existential
quantifier binds the newly introduced monadic predicate, and which 
is 1-satisfiable if and only if the original formula is~(Lemma \ref{th:step1}). 
\item This existential quantifier is afterwards replaced by a universally quantified 
 formula~(Lemma \ref{lem:Hex}). 
\end{enumerate}

\begin{defi} \label{def:SKO}
Let $\parap$ be a new monadic predicate symbol. The operator
$\Psip{\cdot}$, to be applied to prenex formulas from outside to inside,
is defined by
\begin{iteMize}{$\bullet$}
\item $\Psip{\A x A(x)}$ = $\A x\Psip{A(x)}$;
\item $\Psip{\E x A(x)}$ = $\A x(\parap(x) \Impl \Psip{A(f(x,\tup{y}))}$,
     where $f$ is a new (Skolem) function symbol and $\tup{y}$ are the
      free variables in $\E x A(x)$;
\item $\Psip{A}$ = $A$, if $A$ is quantifier free.
\end{iteMize}
The SAT-\emph{Skolem form} $\skp{A}$ of $A$ is obtained by
moving all (universal) quantifiers in $\Psip{A}$ to the front and
inserting one occurrence of $\det$ immediately after the quantifiers.
More precisely,
\[
\skp{A} = \A \tup{x}\det(\Psip{A}^-)
\]
where $\tup{x}$ are the bounded variables in $\Psip{A}$, and $\Psip{A}^-$
denotes $\Psip{A}$ after the removal of all quantifier occurrences.
\end{defi}
Note that applying the operator $\skp{\cdot}$ is not sufficient for
our purpose, since $\skp{P}$ is 1-satisfiable for all formulas $P$. 
For instance, $P = \exists x (\det A(x) \And \neg \det A(x))$ is not
1-satisfiable, while $\skp{P} = \A x (\parap(x) \Impl (\det A(f(x)) 
\And \neg \det A(f(x)))$ is 1-satisfiable.  (A model of the latter
formula is 
obtained by setting $v_\I(\parap)(d)=0$ for all $d$ in the domain~$D$.)
However $\skp{\cdot}$ does preserve 1-(un)satisfiability when 
the  condition $\sup\{v_\I(\parap)(d) \mid d \in D\}=1$
is imposed in addition.
As shown in the following theorem, this amounts to adding the formula
$\E x \parap(x)$ conjunctively to $\skp{P}$.
Henceforth we will slightly widen our focus by considering \emph{conjunctions} of prenex formulas.

\begin{lem}[Step 1]
\label{th:step1}
Let $A_1$, \ldots, $A_m$ be prenex formulas. Then
\[
\bigr({\AND_{1 \le i \le m}}A_i\bigl)\, \in \oSAT 
\ \iff \ 
\bigr(\E x \parap(x) \And {\AND_{1 \le i \le m}\skp{A_i}}\bigl)\, \in \oSAT .
\]
\end{lem}
\proof
Observe that $\skp{A_i}$ is of the form
\[
\A y_1\ldots \A y_k \A x_1\ldots \A x_n \det (\parap(x_1) \Impl \ldots (\parap(x_n)
   \Impl A_i^{sk})\ldots)
\]
where $A_i^{sk}$ denotes the quantifier free part of $A_i$ with
existentially bound variables replaced by Skolem terms 
$f_j(x_j,\tup{y_j})$ ($1 \le j \dots n$), as
specified in Definition~\ref{def:SKO}.

\theifcase
We first show that $$ (\ast) \; \modelsg (\E x \parap(x) \And \skp{A_i}) \Impl
\A y_1\ldots \A y_k \exists v_1 \ldots \exists v_n A_i^{sk}[^{f_j(x_j,\tup{y_j})}/_{f_j(v_j,\tup{y_j})}]$$ 
where $v_j$ are fresh variables and 
$A_i^{sk}[^{f_j(x_j,\tup{y_j})}/_{f_j(v_j,\tup{y_j})}]$ stands for
$A_i^{sk}$ in which the $x_j$ in Skolem terms are replaced by these new variables.

Since
$\modelsg \det\A x B \Iff \A x\det B$ and $\modelsg \det B \Impl B$
for all formulas~$B$ (Lemma \ref{lemma:properties}(1),(4)),
we can remove the indicated occurrence of $\det$ in
$\skp{A_i}$. Then we use
$\modelsg \A x (B^{(x)} \Impl C(x)) \Impl (B^{(x)} \Impl \A x C(x))$ 
(Lemma~\ref{lemma:properties}(5))
to put all universally quantified variables $y_j$ immediately in front of $A_i^{sk}$
and
$\modelsg A(f_j(x_j,\tup{y_j})) \Impl \E v_j A(f_j(v_j,\tup{y_j}))$ (Lemma \ref{lemma:properties}(3)), 
where $v_j$ is a new variable.  Finally by
using Lemma \ref{lemma:properties}(6) we move existential quantifiers 
immediately in front
of all occurrences of~$\parap$ to get
$$ \modelsg \skp{A_i} \Impl (\exists x \parap(x) \Impl \ldots (\exists x \parap(x)
 \Impl \A y_1\ldots \A y_k \exists v_1, \ldots \exists v_n A_i^{sk}[^{f_j(x_j,\tup{y_j})}/_{f_j(v_j,\tup{y_j})}])\ldots)$$
 from which $(\ast)$ follows straightforwardly.
Therefore, if  $(\E x \parap(x) \And \skp{A_i}) \in \oSAT$ then also 
$\A y_1\ldots \A y_k, \exists v_1, \ldots \exists v_n A_i^{sk}[^{f_j(x_j,\tup{y_j})}/_{f_j(v_j,\tup{y_j})}] \in \oSAT$. In fact, it
is easy to transform a model of the latter formula into a model of $A_i$.
%Let $\I$ be a model of the latter formula. An interpretation $\I'$ that is a
%model for $A_i$ is therefore constructed by assigning $\I{A(x_i)} = \I 

\onlyifcase
Suppose that the interpretation $\I$ is a model for $A_i$ for $1\le i \le m$. 
By Proposition~\ref{prop:LS} we can assume that 
$\I$ has a countably infinite domain~$D = \{d_1, d_2, \ldots \}$. 
To obtain a model $\Ip$ with 
the same domain~$D$ for the formula
$\E x \parap(x) \And {\AND_{1 \le i \le m}\skp{A_i}}$  we have to augment $\I$
by a suitable interpretation of $\parap$ and of the Skolem function symbols.
In particular, to achieve $\val{\Ip}{\E x \parap (x)} =1$ we 
assign $v_{\Ip}(\parap)(d_i) = w_i$ in such
a manner that $\sup_i w_i =1$, but $w_i\neq 1$ for all~$i\ge 1$.

In interpreting the Skolem functions we have to make sure that for each step of the
transformation $\Psip{B_i}$, replacing an existential quantifier in $\E x B_i(x)$,
$$\val{\Ip[d/x,\tup{e}/\tup{y}]}{\parap(x) \Impl \Psip{B_i(f(x,\tup{y}))}}=1,
$$
for all $d \in D$ and all $\tup{e} \in D^n$,
where $n$ is the number of free variables in $\E x B_i(x)$. To this aim
we use the assumption that $\val{\I[\tup{e}/\tup{y}]}{\E x B_i(x)} =1$.
This means that for any $d \in D$ there is
a further domain element $d'$ such that
$\val{\Ip[d/x]}{\parap(x)} \le \val{\I[d'/x,\tup{e}/\tup{y}]}{B_i(x)}$.
We assign $v_{\Ip}(f)(d,\tup{e}) = d'$. If there are no more
existential quantifiers in $B_i$ then we are done, as 
$\Psip{B_i} = B_i$ and therefore $\val{\I[d'/x,\tup{e}/\tup{y}]}{B_i(x)}$
= $\val{\Ip[d/x,\tup{e}/\tup{y}]}{\Psip{B_i(f(x,\tup{y}))}}.$
Otherwise we proceed by induction on the number
of existential quantifiers replaced by applying~$\Psipo$, with (essentially)
the presented argument as inductive step.
\qed

We will replace the newly introduced existential quantified formula 
$\E x \parap(x)$ by a conjunction of suitable universal formulas. 
This way we will finally obtain a purely universal formula that 
is 1-satisfiable if and only if the original formula is 1-satisfiable.
To this aim we first introduce a notation that will be useful also in
the next section.

\begin{defi} \label{def:orderrelation}
$A \dle B \defeq \det(A \Impl B)$ and
$A \dl B \defeq \neg\det(B \Impl A)$.
\end{defi}
It is straightforward to check that the suggestive symbols are justified by
\[
\begin{array}{lc@{}l}
\val{\I}{A\,{\dle}B}  =
 \begin{cases}
   1            & \msp \mbox{if } \val{\I}{A} \le \val{\I}{B}\\
   0            & \msp \mbox{otherwise}
\end{cases}
& \ \ \mbox{ and } \  \ &
 \val{\I}{A\,{\dl}B} = 
\begin{cases}
1               &  \msp \mbox{if } \val{\I}{A} < \val{\I}{B}\\
0               &  \msp \mbox{otherwise}. 
\end{cases}
\end{array}
\]

\begin{defi} \label{def:ex_chain}
Let $A$ be a conjunction of prenex formulas of $\Gd$ and
$p_1, \ldots, p_k$ the predicate symbols occurring in $\skp{A}$.
(Note that $\parap \in \{p_1, \ldots, p_k\}$.)
\[
\Hex(\parap,A) \defeq \AND_{1 \le i \le k}
                \A\tup{y_i}(\top \dle p_i(\tup{y_i})
               \Or p_i(\tup{y_i}) \dl \parap(f_{p_i}(\tup{y_i}))),
\]
where $\tup{y_i}$ is a sequence of fresh variables, according to the
arity of $p_i$, and $f_{p_i}$ is a fresh function symbol of corresponding
arity.
\end{defi}

\begin{lem}[Step 2]  
\label{lem:Hex}
Let $A = \AND_{1 \le i \le m}\ A_i$ where 
$A_1$, \ldots, $A_m$ are prenex formulas.  Then
\[
\bigr(\E x \parap(x) \And {\AND_{1 \le i \le m}\skp{A_i}}\bigl)\, \in \oSAT
\ \iff \ 
\bigl(\Hex(\parap,A) \And {{\AND_{1 \le i \le m}}\skp{A_i}}\bigr) \in \oSAT .
\]
\end{lem}

\proof
For the whole proof let 
$p_1, \ldots, p_k$ be the predicate symbols occurring in $\skp{A}$.

\onlyifcase
Let $\I$ be a model of $\E x \parap(x) \And \AND_{1 \le i \le m}\skp{A_i}$
with domain~$D$.
For every $\tup{d} \in D^n$, where $n$ is
the arity of $p_i$ the following holds: either $\val{\I[\tup{d}/\tup{y_i}]}{p_i(\tup{y_i})} = 1$
or $\val{\I[\tup{d}/\tup{y_i}]}{p_i(\tup{y_i})} < 1$. In the former case
the first disjunct $\top \dle p_i(\tup{y_i})$ of the
relevant conjunct in  $\Hex(\parap,A)$ evaluates to~$1$.
In the latter case, since we have $\val{\I}{\E x \parap(x)} =1$,
we can  extend $\I$ by a valuation function for
the new function symbols $f_{p_i}$ in such a manner that
$\val{\I[\tup{d}/\tup{y_i}]}{\parap(f_{p_i}(\tup{y_i}))}$ $>$
$\val{\I[\tup{d}/\tup{y_i}]}{p_i(\tup{y_i})}$ holds. But this implies
that the second disjunct in  $\Hex(\parap,A)$ is evaluated to~$1$.

\theifcase
Let $\I$ be a model of $\Hex(\parap,A) \And {\AND_{1 \le i \le m}\skp{A_i}}$
with domain~$D$.
If $\val{\I}{\E x\parap(x)}=1$ then we are done.
Otherwise, $\val{\I[d/x]}{\parap(x)}<1$ for all $d \in D$. 
Note that $\parap \in \{p_1, \ldots, p_k\}$ and therefore 
$\val{\I}{\Hex(\parap,A)}=1$ implies that 
$\val{\I[d/x]}{\parap(x)}<\val{\I[d/x]}{\parap(f_\parap(x))}$,
for every $d \in D$,
%we have 
% $\val{\I[x/d]}{\parap(x)}$ $<$ $\val{\I[x/d]}{\parap(f_\parap(x))}$,
since $\val{\I[d/x]}{\top \dle \parap(x)} < 1$. Consequently 
$\sup_{d \in D}\val{\I[d/x]}{\parap(x)}= v$ for some
$v<1$, but nevertheless
$\val{\I[d/x]}{\parap(x)}\neq v$ for all $d \in D$.
$\val{\I}{\Hex(\parap,A)}=1$ also implies
that for every $\tup{d} \in D^n$, where $n$ is
the arity of $p_i$, we have either
$\val{\I[\tup{d}/\tup{y_i}]}{p_i(\tup{y_i})}=1$ or
$\val{\I[\tup{d}/\tup{y_i}]}{p_i(\tup{y_i})}<v$. In other words:
no atomic formula is assigned a value in the interval $[v,1)$ by~$\I$.
We may therefore define a new interpretation $\Ip$ over
the same domain $D$ by setting
$\val{\Ip[\tup{d}/\tup{y_i}]}{p_i(\tup{y_i})} = 
\val{\I[\tup{d}/\tup{y_i}]}{p_i(\tup{y_i})} + (1-v)$, whenever
$\val{\I[\tup{d}/\tup{y_i}]}{p_i(\tup{y_i})} \neq 1$ and 
$\val{\I[\tup{d}/\tup{y_i}]}{p_i(\tup{y_i})} \neq 0$. Otherwise the
corresponding truth value remains the same, i.e., ~$1$ or~$0$, respectively.

It remains to show that $\Ip$ is a model of $\E x \parap(x) \And {\AND_{1 \le i \le m}\skp{A_i}}$.  
By definition of~$\Ip$, $\sup_{d \in D}\val{\Ip[d/x]}{\parap(x)}= 1$.
To complete the argument remember that each
$\skp{A_i}$ is of the form $\A \tup{x}\det B^-_i$ where 
$B^-_i$ is
 $(\parap(x_1) \Impl \ldots (\parap(x_n) \Impl A_i^{sk})\ldots)$. 
Therefore $\val{\I}{\skp{A_i}} = 1$ implies that 
$\val{\I[\tup{d}/\tup{x}]}{B^-_i}=1$ for 
every appropriate tuple $\tup{d}$ of domain elements. 
This means that the evaluation
reduces to that of a quantifier free formula of~$\Gd$.
Now recall from  Lemma~\ref{lemma:prop} that
whether a given interpretation~$\I$ satisfies
a quantifier free formula only depends on the relative order 
of assigned truth values
below~$1$ and above~$0$, but not on their absolute values. 
Therefore, just like~$\I$,
also $\Ip$ is a model of $\skp{A_i}$ for $1 \le i \le m$.
\eop

\begin{exa}
Let $F = \E x p(x) \And \A y \neg \det p(y)$.\\
Step 1: by Lemma \ref{th:step1}, 
$$F \in \oSAT \quad \mbox{if and only if} \quad \exists x \parap(x)
 \And \skp{\E x p(x)} \And
\A y \neg \det p(y) \in \oSAT,$$ where 
$\skp{\E x p(x)} = \forall x (q(x) \to p(f(x)))$.\\
Step 2: the existential quantifier is removed by translating 
$\exists x \parap(x)$
into $\Hex(\parap,F)$:
$$\begin{array}{rl}
 & \forall y_1 (\top \dle \parap(y_1) \Or \parap(y_1) \dl \parap(f_\parap(y_1)))\\
 \And &
   \forall y_2 (\top \dle p(y_2) \vee p(y_2) \dl \parap(f_p(y_2)))\\
\And & \forall x (q(x) \to p(f(x))) \ .
\end{array}
$$
According to Lemma~\ref{lem:Hex} $\Hex(\parap,F) \And \skp{\E x p(x)} 
\And \A y \neg \det p(y)$
is 1-satisfiable if and only if $F$ is 1-satisfiable.
We refer to it as the Skolemized form of $F$ with respect to satisfiability.
\end{exa}

Although standard Skolemization does not preserve $1$-satisfiability 
for all prenex formulas of~$\Gd$, it does so for formulas in which the 
quantifier free part is preceded by~$\det$. 
Indeed, formulas of the form
$\Q \tup{x}\det B$, where $B$ is quantifier-free, 
can be Skolemized in the standard way,
i.e., every existentially quantified variable $x$ is replaced by a
Skolem term~$f(\tup{y})$, where $\tup{y}$ denotes the variables
bound by universal quantifiers in the scope of which $x$ occurs (perfectly dual to
Definition \ref{def:Skolem}).
We will denote by $\forall \overline{x}\det B^{S}$ the formula arising from $\Q \tup{x}\det B$ 
in this manner.

\begin{lem}
\label{lemma:SKOsat}
$\Q \tup{x}\det B \in \oSAT \ \iff \ \forall \overline{x}B^{S} \in \oSAT$.
\end{lem}

\proof
\theifcase Easy. For
\onlyifcase note that a formula $\exists y \det B'(y)$ evaluates to $1$ under an interpretation
$\I$ if and only if $\val{\I[d/y]}{B'} = 1$ for some domain element $d$.
\eop

This observation can be exploited to achieve a more efficient translation 
of conjunctions of prenex formulas, as stated in the following corollary.
\begin{cor}
\label{cor:SKOsat_improved}
Let $F = {\AND_{1 \le i \le m}}A_i$, where $A_i$ are prenex formulas of $\Gd$
and for $1 \le i \le m_1$ ($m_1 \le m$) $A_i$ is of the form
$\Q \tup{x}\det B_i$ for some quantifier free formula $B_i$.  Then
\[ F \, \in \oSAT
\ \iff \
\bigl(\AND_{1 \le i \le m_1} \forall \overline{x}A_i^{S} \wedge 
\Hex(\parap, \AND_{m_1 \le i \le m} A_i) \And {{\AND_{m_1 \le i \le m}}\skp{A_i}}\bigr) \in \oSAT \ .
\]
\end{cor}

%****************************************************************
%****************************************************************
\section{Theorem Proving} \label{sec:TP}
%******************************************************************
%****************************************************************

Let $F$ be a conjunction of prenex formulas.
The results of the last section amount to the following central 
``preprocessing steps'' for 
automated theorem proving: 
\begin{iteMize}{$\bullet$}
\item Testing validity of $F$ can be reduced to testing validity of a purely
      existential formula~$\E \tup{x} G$ (Corollary \ref{cor:SKOval}).
\item Testing 1-satisfiability of $F$ can be reduced to testing 1-satisfiability
      of a purely universal formula~$\A \tup{x} G'$ (Corollary \ref{cor:SKOsat_improved}).
\end{iteMize}
Note that the first problem is $\Sigma_1$-complete (because $\Gd$ is recursively
axiomatizable~\cite{hajek1}), whereas the second problem
is $\Pi_1$-complete~\cite{hajek2}, just like the corresponding problems for 
classical logic.
However, in contrast to classical logic, the problems are not simply dual to
each other:  to obtain $G'$ we even had to extend the signature of 
$F$ and $G$ by introducing a new predicate symbol. 
Nevertheless we can treat $\E \tup{x} G$ and $\A \tup{x} G'$ in the
same manner for our next step towards efficient theorem proving: translating
the quantifier free part ($G$, $G'$) into a suitable normal form.

In our case this normal form will directly correspond to so-called
order clauses that refer to the (classical) theory of dense linear orders
with endpoints. In this manner both, the validity and the 1-satisfiability
problem for prenex~$\Gd$, are reduced to detecting the (un)-satisfiability
of specific sets of order clauses. For handling the latter problem we can
rely on results from the literature on automated theorem proving
using ordered chaining resolution, 
as we will point out in Section~\ref{ss:resolution}. 

In fact, a particular normal form for propositional formulas of~$\Gd$,
called {\em chain normal form},
has already been described in the literature, see e.g.,~\cite{BCF01,BF10}. To recall this
notion let us use, in addition to the abreviations $\dl$ and $\dle$ 
(Definition~\ref{def:orderrelation}), also $A \deq B$ as an abbreviation for 
$\det(A \Iff B)$. 
Clearly $\val{\I}{A\deq B} =1$ iff $\val{\I}{A} = \val{\I}{B}$.

\begin{defi} 
Let $F$ be a quantifier-free formula of $\Gd$ and let $A_1, \dots ,A_n$ be the atoms
occurring in~$F$. A $\det$-{\em chain} over $F$ is a formula of the form
\begin{equation*} (\bot \Join_0 A_{\pi(1)}) \And (A_{\pi(1)} \Join_1 A_{\pi(2)})
\And  \cdots \And (A_{\pi(n-1)} \Join_{n-1} A_{\pi(n)}) \And (A_{\pi(n)}
\Join_n \top) 
\end{equation*}
where $\pi$ is a permutation of $\{1, \dots, n\}$, $\Join_i$ is either
$\dl$ or $\deq$, but at least one of the $\Join_i$'s stands for $\dl$.

By $\Ch{F}$ we denote the set of all  $\det$-chains over~$F$.
\end{defi}

The following follows immediately from Theorem~17 of~\cite{BCF01}.

\begin{thm} \label{th:ch}
Let $F$ be of the form $\AND_{1\le i \le n} \A \tup{x_i} \det F_i$,
where $F_i$ is quantifier free. Then there exist $\Gamma_i
\subseteq \Ch{F_i}$ for all $1 \le i \le n$ such that
$$\modelsg F \; \ifff {\AND_{1\le i \le n}}{\OR_{\,\,\,\,C \in \Gamma_i}}
                  \hspace*{-1ex} C.$$
\end{thm} 

While Theorem~\ref{th:ch} can be used, in principle, to translate Skolemized
formulas into a kind of disjunction normal form, the translation as well as
the resulting normal form is excessively complex in general. 
To appreciate the problem, note that
$\Ch{F}$ contains a super-exponential number of different $\det$-chains (with
respect to the length of $F$) in general. Clearly we need an alternative
translation to normal form to obtain a practically feasible proof method. 
A suitable normal form is presented below.

%%%%%%%%%%%%%%%%%%
\subsection{Structural Translation to Order Clauses} \label{ss:clauseform}
%%%%%%%%%%%%%%%%%

It is well known from classical logic that the combinatorial explosion that
may arise in any language preserving translation of arbitrary 
complex formulas into conjunctive normal form can be avoided by a
{\em structural translation}. The latter introduces new predicate symbols to define
appropriate abbreviations of subformulas, see~\cite{PG,Handbook}.
Our translation of Skolemized $\Gd$-formulas 
to clausal form proceeds in an analogous manner. We consider this as a two-step
process that can roughly be described as follows:
\begin{enumerate}[(1)]
\item The quantifier free part is efficiently reduced to a formula of an extended 
      language, involving a conjunction of simple equivalences that introduce
      new predicate symbols as abbreviations for subformulas (``definitional normal form'').
\item The resulting $\Gd$-formula 
      is translated into a set of clauses, where
      the literals are of the form $s < t$ or $s \le t$, referring to the
      (classical) theory of dense total orders with endpoints (``definitional clause form'').
\end{enumerate}
For step~1 we introduce the following notions. (Remember that negations $\neg F$
appear as $F\Impl\f$.)

\begin{defi} \label{def:dfequiv}
For any quantifier free formula $F$ of the form $F_1 \circ F_2$, where 
$\circ \in\{\And,\Or,\Impl\}$, let
$$ \df{F} \defeq [p_F(\tup{x}) \deq 
           (p_{F_1}(\tup{x_1}) \circ p_{F_2}(\tup{x_2}))]$$
where $p_F, p_{F_1}, p_{F_2}$ are new predicate symbols and 
$\tup{x}, \tup{x_1}, \tup{x_2}$ are the tuples of variables occurring in 
$F, F_1, F_2$, respectively. If $F$ is of the form $\det F_1$ then
$$ \df{F} \defeq [p_F(\tup{x}) \deq \det p_{F_1}(\tup{x_1})].$$
If $F$ is atomic then $p_F(\tup{x})$ is simply an alternative 
denotation for~$F$.
\end{defi}

Depending on whether we are interested in 1-satisfiability or in 
validity we need two different normal forms based on the
equivalences introduced in Definition~\ref{def:dfequiv}.

\begin{defi} \label{def:df}
Let $A$ be a quantifier free formula.
The {\em definitional normal form for 1-satisfiability} is
defined as
\[
\DEFsat{A} \defeq \det p_A(\tup{x})
        \And \bigl(\hspace*{-1ex}\AND_{F \in \subf{A}}\hspace*{-1ex}\df{F}\bigr)
\]
The {\em definitional normal form for validity} is
defined as
\[
\DEFval{A} \defeq 
    \bigl(\hspace*{-1ex}\AND_{F \in \subf{A}}\hspace*{-1ex}\df{F}\bigr)
           \Impl \det p_A(\tup{x})
\]
In both cases, $\subf{A}$ denotes the set of all non-atomic
subformulas of $A$, $\tup{x}$ is  the tuple of variables occurring in
$A$, and $p_A$ is a new predicate symbol.
\end{defi}
To prove the soundness of the definitional normal form for validity
for existential formulas the following lemma is needed. Its proof
requires Herbrand's Theorem. 

\begin{lem} \label{lemma:delta_ex}
For all quantifier free formulas $A$ of $\Gd$:
$\modelsg \E\tup{x}A \IFF \modelsg \E\tup{x}\det A$.
\end{lem}
 
\proof
The direction from right to left is trivial. 
The other direction is obtained as follows:
\[
\begin{array}{rl@{\quad}l}
     & \modelsg \E\tup{x}A & \\%[1.2ex]
\IMPLIES & \modelsg\OR_{i=1}^{n} A(\tup{t_i}) \mbox{ for appropriate }
               \tup{t_1}, \ldots, \tup{t_n}
                  &  \mbox{ by Theorem \ref{univH} (Herbrand Theorem)} \\%[1.2ex]
\IMPLIES & \modelsg \det \OR_{i=1}^{n} A(\tup{t_i}) 
         &  \mbox{ since } \modelsg F \mbox{ iff } \modelsg \det F \\%[1.2ex]
\IMPLIES & \modelsg \OR_{i=1}^{n} \det A(\tup{t_i}) 
           &  \mbox{ by Lemma~\ref{lemma:properties}(2) }\\
\IMPLIES & \modelsg \E\tup{x}\det A(\tup{x}) & 
\mbox{ by laws of $\Gd$.} 
\end{array} 
\]
Note that in the second and in the last step we used the fact that for quantifier
free $A(\tup{x})$ the formula $\E\tup{x}A(\tup{x})$ is already in Skolem form. 
\qed

We remark that the above lemma does not hold when $A$ is not quantifier free.
%%%%%  sat/valid <=> DEFform sat/val
\begin{lem} \label{lemma:df}
Let $A$ be a quantifier free formula with free variables~$\tup{x}$.
\begin{enumerate}[\em(a)]
\item $\A\tup{x}A$ is 1-satisfiable iff $\A\tup{x}\DEFsat{A}$ is 1-satisfiable.
\item $\E\tup{x}A$ is valid iff $\E\tup{x}\DEFval{A}$ is valid.
\end{enumerate}
\end{lem}
\proof
Note that all the relevant subformulas of $\DEFsat{A}$ and $\DEFval{A}$
are preceded by an occurrence of~$\det$. Thus we obtain in the
same manner as for the definitional normal forms of classical logic 
(see \cite{PG,Handbook}) that 
$\vI(\df{F(\tup{x})})=1$ iff $\vI(F(\tup{x}))=\vI(p_F(\tup{x}))$ for
all subformulas $F$ of~$A$. Consequently we have:
\begin{enumerate}[($\ast_1$)]
\item Every model of $A$ can be extended to a model of
      $\DEFsat{A}$; conversely every model of $\DEFsat{A}$ is also a model
      of~$A$.
\item All extensions  of a model of $A$ to the language that
       additionally contains the new predicate symbols $p_F$ for $F \in \subf{A}$
       are models of $\DEFval{A}$; conversely every model $\I$ of 
       $\DEFval{A}$ where $\vI(p_A(\tup{x}))=1$  is also a model
       of~$A$.
\end{enumerate}

\noindent To obtain (a) from ($\ast_1$) it suffices to remember that, for arbitrary
formulas~$F$, $\A x F$ is 1-satisfiable iff $F$ there is an interpretation~$\I$
such that $\I[\tup{d}/\tup{x}]$ is a model of~$F$
for every variable assignment $[\tup{d}/\tup{x}]$.

To obtain (b) something more is needed, since in general an interpretation 
$v_{\I}$ can be a model for $\E x F$ even if $\val{\I[d/x]}{F(x)} < 1$
for each variable assignment. 
However note that this cannot happen if $F$ is of the form $\det G$.
More precisely:
\begin{enumerate}[($\ast_1$)]
\item[($\ast_3$)] For all fomulas~$G$, $\modelsg \E\tup{x} \det G$ 
     iff for every interpretation $\I$
     there is a variable assignment $[\tup{d}/\tup{x}]$ such that 
      $\val{\I[\tup{d}/\tup{x}]}{G(\tup{x})} = 1$.
\end{enumerate}
Therefore, using Lemma~\ref{lemma:delta_ex}, we argue as follows:
\[
\begin{array}{rl@{\quad}l}
     & \modelsg \E\tup{x}A & \\%[1.2ex]
\IFF & \modelsg \E\tup{x}\det A  & 
           \mbox{ by Lemma~\ref{lemma:delta_ex}} \\
\IFF & \modelsg \E\tup{x}\det\DEFval{A} & 
          \mbox{ by } (\ast_2) \mbox{ and } (\ast_3)\\
\IFF &  \modelsg \E\tup{x}\DEFval{A} & 
           \mbox{ by Lemma~\ref{lemma:delta_ex}}\rlap{\hbox to 121
             pt{\hfill\qEd}}
\end{array}\medskip
\]

\noindent We now switch to step 2 of the translation into clausal form, which 
results in a ``logic free'' syntax by considering all
predicate symbols as function symbols
and the special atomic formulas $\top$ and $\f$ as constant symbols. 
More precisely,
an atomic $\Gd$-formula like $p(x,f(x,y))$  will no longer be
considered to be a \emph{formula} of the new language, 
but now simply appears as a \emph{term} containing two binary function symbols.
We will use $\Terms$ to denote the set of all 
terms arising in this manner.

\begin{defi} \label{def:strcf}
An \emph{order literal} is an expression of the form
$s < t$ or $s\le t$, where $s, t \in \Terms$.
An \emph{order clause} is a finite set of literals,
representing  a disjunction of its elements.%
\footnote{
In \cite{BGcade} order
clauses are defined as \emph{multisets}. However it follows from
results in \cite{BGcade} concerning redundancy that we may
alternatively define clauses as sets.
}
\end{defi}
 
Semantically order clauses refer to the following classical structure.

\begin{defi}
By a {\em dense total order} $\O$ we mean an interpretation of 
the predicate symbols $<$ and $\le$,  taking terms in $\Terms$ as arguments,
where $<$ refers to a strict and dense total (linear) order
and $\le$ is interpreted 
as the reflexive closure of~$<$. If, in addition, the {\em endpoint} axioms 
$\A x(\f \le x)$, $\A x(x\le \top)$, and $\f<\top$
are satisfied we call $\O$ a \DTOE-model. 

A set of order clauses $\CS$ is {\em \DTOE-satisfiable} if the conjunction
of elements of $\CS$ has a dense total order with endpoints $\f$ and $\top$
as a model; otherwise $\CS$ is called {\em \DTOE-unsatisfiable}.
\end{defi}

The normal form that will be used in the next section is a translation of 
the definitional normal forms in Definition \ref{def:df} into suitable order clauses.
\begin{defi} \label{def:defc}
Let $A$, $B$, and $C$ be atomic formulas.
\[
\begin{array}{lcl}
\cl{C \deq (A \And B)} & \defeq &
      \{\{C \le A\},\ \{C \le B\},\ \{A\le C, B\le C\} \}\\
\cl{C \deq (A \Or B)} & \defeq &
      \{\{A \le C\},\ \{B \le C\},\ \{C\le A, C\le B\} \}\\
\cl{C \deq (A \Impl B)} & \defeq &
      \{\{A \le B, C\le B\},\ \{\top \le C, B < A\}, \\
      &&\phantom{\{}   \{\top \le C, C \le B\},\ 
        \{B \le C\}\} \\
\cl{C \deq \det A}  & \defeq &
      \{ \{C \le \bot, \top \le A\},\ \{\top \le C, A < \top \}\} 
\end{array}
\]
For a quantifier free formula $G$ the \emph{definitional clause form for 
1-satisfiability}
is defined as
\[
\CLsat{G} \defeq \{\{\top \le p_G(\tup{x})\}\} \cup \bigcup_{F \in \subf{G}}
                          \hspace*{-1ex}\cl{\df{F}}\bigr)
\]
and the \emph{definitional clause form for validity}
is defined as
\[
\CLval{G} \defeq \{\{p_G(\tup{x}) < \top \}\} \cup \bigcup_{F \in \subf{G}}
                          \hspace*{-1ex}\cl{\df{F}}\bigr),
\]
where $\subf{G}$ denotes the set of all non-atomic
subformulas of $G$, $\tup{x}$ is the tuple of variables occurring in
$G$, and $p_G$ is a new predicate symbol.
\end{defi}

%%%%%%%%%%% DEFform sat/valid <=> Clauseform sat/unsat %%%%%%%%%
\begin{lem}  \label{lemma:defcf}
Let $A$ be a quantifier free formula with free variables~$\tup{x}$.
\begin{enumerate}[(a)]
\item $\A\tup{x}\DEFsat{A}$ is 1-satisfiable iff 
     $\CLsat{A}$ is \DTOE-satisfiable. 
\item $\E\tup{x}\DEFval{A}$ is valid iff $\CLval{A}$ is 
     \DTOE-unsatisfiable. 
\end{enumerate}
\end{lem}

\proof
We have to check that the clauses specified in Definition~\ref{def:defc} 
are equivalent to the corresponding subformulas involving `$\deq$' 
in the definitional forms specified
in Definition~\ref{def:df}.

\begin{iteMize}{$\bullet$}
\item $C \deq (A \Impl B)$: it is not difficult to see that the following formulas are equivalent in $\Gd$,
i.e. for each interpretation $\I$ we have 
\[ 
\begin{array}{lcl}
\val{\I}{\det(C \Iff (A \Impl B))} &=&
       \val{\I}{(A \dle B \And \det C) \Or (B \dl A \And C\dle B \And  B\dle C)}.
\end{array}
\]
By applying the law of distribution to the formula at the right hand side 
we obtain 
the conjunction of the following six formulas (note that we can express $\det C$ by the equivalent formula $\top \dle C$, cf.\ Definition~\ref{def:orderrelation}):
  \[ 
\begin{array}{r@{ \ \ }c@{ \ \ }l@{\ \qquad}l}
 A \dle B &\Or& B \dl A & (1) \\
 A \dle B &\Or& C \dle B & (2) \\
 A \dle B &\Or& B \dle C & (3) \\
 \top \dle C &\Or& B \dl A & (4)\\
 \top \dle C & \Or & C\dle B & (5) \\
 \top \dle C &\Or& B \dle C & (6) 
\end{array}
\] 
Note that conjunct $(1)$ is valid and that $B \dle C$ is entailed by~(6).
$B \dle C$ in turn entails conjuncts $(3)$ and~$(6)$. 
Thus we obtain the following
 four conjuncts that directly correspond to $\cl{C \deq (A \Impl B)}$:
  \[ 
\begin{array}{r@{ \ \ }c@{ \ \ }l@{\ \qquad}l}
 A \dle B &\Or& C \dle B &  \\
 \top \dle C &\Or& B \dl A & \\
 \top \dle C &\Or& C \dle B & \\
 B \dle C &&&
\end{array}
\] 
\item $C \deq (A \And B)$: $\det(C \Iff (A \And B))$ is easily seen
   to be equivalent to the conjunction of
\[ 
\begin{array}{r@{ \ \ }c@{ \ \ }l@{\ \qquad}l}
 C \dle A &&&  \\
 C \dle B &&&\\
 A \dle C &\Or& B \dle C 
\end{array}
\]
that directly correspond to $\cl{C \deq (A \And B)}$.
\item $C \deq (A \Or B)$: $\det(C \Iff (A \Or B))$ is 
   equivalent to the conjunction of 
\[ 
\begin{array}{r@{ \ \ }c@{ \ \ }l@{\ \qquad}l}
 A \dle C &&&  \\
 B \dle C &&&  \\
 C \dle A &\Or& C \dle B 
\end{array}
\]
that directly correspond to $\cl{C \deq (A \Or B)}$.
\item $C \deq \det A$:  $\det (C \Iff \det A)$ is equivalent to the 
 conjunctions of the following two disjunctions
\[ 
\begin{array}{r@{ \ \ }c@{ \ \ }l@{\ \qquad}l}
C \dle \bot &\Or& \top \dle A &\\
\top \dle C&\Or& A \dl \top
\end{array}
\]
that directly correspond to $\cl{C \deq \det A}$. 
\end{iteMize}
So far we have argued about equivalences 
within $\Gd$. But note that formulas
of the form $A \dle B$ or $A\dl B$ evaluate to either $0$ or~$1$
in every interpretation. Therefore disjunction and conjunction 
reduce to their classical counterparts and we can directly translate 
$A \dle B$ and $A\dl B$ into order literals $A\le B$ and $A < B$,
respectively. In this manner we obtain sets of order clauses that
are \DTOE-satisfiable iff the corresponding $\Gd$-formulas
are 1-satisfiable. 
In the case of  $\CLsat{A}$  the clause $\{\top \le p_G(\tup{x})\}$
directly codes the claim that $A \in \oSAT$, whereas for $\CLval{A}$ the
clause $\{p_G(\tup{x}) < 1\}$ ensures that $A$ is valid iff 
$\CLval{A}$ is $\DTOE$-unsatisfiable.
\qed

\begin{rem}
A somewhat different structural clause form has been
described in~\cite{BCF01}. Here, following~\cite{BF10}, we have eliminated
a number of redundancies from the originally described sets of order clauses.
\end{rem}

The following theorem combines the various steps of the current and
the last section and points out the efficiency of the overall translation.
%%%%%%%% Summarizing 
%%%  {lemma:df} {lemma:clf} {cor:SKOval}  {lemma:SKOsat}  {cor:SKOsat_improved}

\begin{thm} \label{th:main}
Let $A = {\AND_{1 \le i \le m}}A_i$ where $A_i$ are prenex formulas of $\Gd$.
Then one can construct in polynomial time 
sets of order clauses $\CSsat{A}$ and $\CSval{A_i}$, where $1 \le i \le n$,
such that
\begin{enumerate}[\em(a)]
\item $A \in \oSAT$ iff $\CSsat{A}$
       is \DTOE-satisfiable.
\item $\modelsg A$ iff $\CSval{A_i}$
          is  \DTOE-unsatisfiable for each $i \in \{1, \ldots, n\}$.
\end{enumerate}
\end{thm}

\proof
(a) \ By Corollary \ref{cor:SKOsat_improved} we obtain a formula
  $B =  {\AND_{1 \le i \le n}}\A\tup{x}_i B_i$, where the $B_i$ are
  quantifier free, such that $A \in \oSAT$ iff $B \in \oSAT$. Remember that
  the conjunction of formulas corresponds to the union of sets of clauses.
  Therefore the combination of Lemmas~\ref{lemma:df} and~\ref{lemma:defcf}
  implies that $A$ is 1-satisfiable iff 
   $\CSsat{A} = \bigcup_{1 \le i \le n} \CLsat{B_i}$ is \DTOE-satisfiable.

(b) \ By Corollary~\ref{cor:SKOval} we obtain a formula
  $B =  {\AND_{1 \le i \le n}}\E\tup{x}_i B_i$, where the $B_i$ are
  quantifier free, such that $\modelsg A $ iff $\modelsg B$. Since a conjunction
  is valid iff every conjunct is valid, 
  Lemmas~\ref{lemma:df} and~\ref{lemma:defcf} reduce the problem
   of checking whether
  $\modelsg A$ to checking whether for each $i \in \{1, \ldots, n\}$
  the clause set $\CSval{A_i}=\CLval{B_i}$ is \DTOE-unsatisfiable.

It finally remains to observe that $\CSsat{A}$ and the $\CSval{A_i}$ are
of polynomial size with respect to the size of~$A$. In particular note
that $\Hex(\parap,A)$ (Definition~\ref{def:ex_chain}) and $\skp{A_i}$ 
(Definition~\ref{def:SKO})  increase the overall size of the formula
only by a linear number of symbols. Also the definitional clause forms
(Definition~\ref{def:defc}) 
are linear in the size of~$A$. Consequently, all mentioned transformations 
can clearly be done in polynomial time. 
\qed

\begin{exa} \label{ex:delta}
We claim that the following formula is valid in~$\Gd$:
$$A = \E x\A y (\det p(y) \Impl p(x))\ .$$
While simple, this example is nevertheless of some interest. 
In particular note that removing the occurrence of~$\det$ in~$A$
results in a formula that is not any longer valid in~$\Gd$, 
although it is classically valid.

According to Section~\ref{ss:skolemval} we obtain the Skolemized form of $A$ as
$$\E x (\det p(f(x)) \Impl p(x))\ .$$
To compute the definitional normal form we have
to introduce the following two ``definitions'' of subformulas as described
in Definition~\ref{def:dfequiv}:
\begin{desCription}
\item\noindent{\hskip-12 pt\bf $\df{\det p(f(x)}$:}\ $p_1(x) \deq \det p(f(x))$,
\item\noindent{\hskip-12 pt\bf $\df{p_1(x) \Impl p(x)}$:}\ $p_2(x) \deq (p_1(x) \Impl p(x))$.
\end{desCription}
The corresponding order clauses according to Definition~\ref{def:defc}
are as follows:
\begin{desCription}
\item\noindent{\hskip-12 pt\bf $\cl{p_1(x) \deq \det p(f(x)}$:}\ ~\\
    $1: \{p_1(x) \le \bot, \top \le p(f(x))\}$\\
    $2: \{\top \le p_1(x), p(f(x)) < \top\}$ 
\item\noindent{\hskip-12 pt\bf $\cl{p_2(x) \deq (p_1(x) \Impl
    p(x)}$:}\ ~\\
    $3: \{ p_1(x)\le p(x), p_2(x)\le p(x)\}$\\
    $4: \{\top\le p_2(x), p(x)<p_1(x) \}$\\
    $5: \{ \top\le p_2(x), p_2(x)\le p(x)\}$\\
    $6: \{ p(x)\le p_2(x)\}$
\end{desCription}
Since we are interested in validity we have to add\\
$\hspace*{8ex} 7: \{p_2(x) < \top \}$ \\
to obtain $\CLval{\det p(f(x)) \Impl p(x)}$ as specified in 
Definition~\ref{def:defc}. 

We will continue this example in Section~\ref{ss:resolution} to illustrate
a machine oriented proof of the \DTOE-unsatisfiability of 
$\CLval{\det p(f(x)) \Impl p(x)}$.
\end{exa}

%%%%%%%%%%%%%%%%%%%%%%%
\subsection{Ordered Chaining Resolution}  \label{ss:resolution}

The results of the previous sections, as summarized in Theorem~\ref{th:main},
reduce the validity as well as the 1-satisfiability problem for prenex $\Gd$
to checking \DTOE-(un)satisfiability of certain 
sets of {order clauses}. 
Fortunately, efficient theorem proving for various types of
order clauses has already received considerable
attention in the literature; see \cite{BGjacm,BGcade}
and the references given there. We finally just extract from this literature
what is needed in our specific case.

We recall some basic notions from automated deduction (see, e.g., \cite{leitsch}).
In particular we identify a {\em substitution} $\sigma$  with a set 
$\{x_1 \gets t_1, \ldots, x_n \gets t_n\}$ and define
$\codom(\sigma) = \{t_1, \ldots, t_n\}$. $E\sigma$ denotes
the result of applying $\sigma$
to an expression $E$, i.e. $E\sigma$ is obtained by replacing for each
$i\in\{1,\ldots,n\}$ all
occurrences of the variable $x_i$ in $E$ by the term $t_i$. Finally, 
recall that a substitution $\sigma$ is called
the {\em most general unifier (mgu)} of terms $s_1, \ldots, s_n$ if
$s_1\sigma = \ldots =s_n\sigma$ and if in addition for all other 
substitutions~$\rho$ where $s_1\rho = \ldots =s_n\rho$ we have
$s_i\rho = (s_i\sigma)\tau$ for some substitution~$\tau$.

We consider the following rules (cf.~\cite{BGcade}) for order clauses:
\bd
\item\noindent{\hskip-12 pt\bf Irreflexivity Resolution:}\
$$\infer{C\sigma}{C \cup \{s<t\}}$$
     where $\sigma$ is the mgu of $s$ and $t$
\item\noindent{\hskip-12 pt\bf (Factorized) Chaining:}\ 
$$\infer{C\sigma \cup D\sigma 
   \cup\{u_i\sigma \sm_{i,j} r_j\sigma \mid 1\le i\le m, 1\le n \}}%
  {C \cup \{u_1\sm_1 s_1, \ldots,u_m\sm_m s_m\} & 
   D \cup \{t_1\sm'_1 r_1, \ldots, t_n\sm'_n r_n\}
  }
$$
  where $\sigma$ is the mgu of $s_1,\ldots,s_m,t_1,\dots,t_n$ and
  $\sm_{i,j}$ is $<$ if and only if either $\sm_i$ is $<$ or $\sm'_j$ is $<$.
  Moreover, $t_1\sigma$ occurs in $D\sigma$ only in inequalities 
  $r`\sm t_1\sigma$.
\ed
These two rules constitute a refutationally complete inference system
for the theory of all total orders in presence of set $\EEf$ of
clauses
$$\{x_i<y_i, y_i<x_i \mid 1 \le i \le n\} 
        \cup \{f(x_1,\ldots,x_n) \le f(y_1,\ldots,y_n)\},$$
where $f$ ranges over 
the set $\FS$ of function symbols of the signature.
Observe that, in translating a formula $P$ from prenex $\Gd$ into
a set of order clauses $\CL{P}$, we treat the predicate symbols of $P$
as function symbols. Additional function symbols occur from Skolemization.

The inference system is not yet sufficiently restrictive for efficient
proof search. We follow
\cite{BGcade} and add conditions to the rules that refer to some
complete reduction order $\succ$ (on the set of all terms).
We write $s \not\succeq t$ if $\neg (s \succ t)$ and $s\neq t$; and
 ``$t$ {\em is basic in (clause)} $C$'' if $t\sm s \in C$ or 
$s\sm t \in C$.
\bd
\item\noindent{\hskip-12 pt\bf Maximality Condition for Irreflexivity
  Resolution:}\ 
  $s\sigma$ is a maximal term in $C\sigma$.
\item\noindent{\hskip-12 pt\bf Maximality Condition for Chaining:}\ 
\begin{enumerate}[(1)]
  \item $u_i\sigma \not\succeq s_1\sigma$ for all $1 \le i \le n$,
  \item $v_i\sigma \not\succeq t_1\sigma$ for all $1 \le i \le m$,
  \item $u\sigma \not\succeq s_1\sigma$ for all terms $u$ that 
       are basic in $C$, and
  \item $v\sigma \not\succeq t_1\sigma$ for all terms $v$ that are
    basic in $D$.
  \end{enumerate}
\ed
For our purposes it is convenient to view the resulting
inference system \MC{} as a set operator.

\begin{defi}
$\MC(\CS)$ is the set of all conclusions of Irreflexivity Resolution
or Maximal Chaining where the premises are (variable renamed copies of)
members of the set of clauses~$\CS$. Moreover, 
$\MC^0(\CS)= \CS$, $\MC^{i+1}(\CS) = \MC(\MC^i(\CS)) \cup \MC^i(\CS)$,
and $\MC^*(\CS) = \bigcup_{i\ge 0} \MC^i(\CS)$.
\end{defi}

The set consisting of the three clauses $\{\bot \le y\}$, $\{y\le \top\}$,
and $\{\bot<\top\}$, corresponding to the endpoint axioms, is called~$\EP$.
The set consisting of $\{y\le x, d(x,y) < y\}$ and
$\{y\le x, x <d(x,y)\}$, corresponding to the usual density axiom,
is called~$\DO$.

The following completeness theorem follows directly from  Theorem 2 of \cite{BGcade}.
%(See also \cite{BGjacm}).

\begin{thm} \label{th:MC}
$\CS$ has a dense total order with endpoints $0$ and $1$
as a model if and only if $\MC^*(\CS \cup \EEf \cup \EP \cup \DO)$
does not contain the empty clause.
\end{thm}

\begin{rem}
Even more refined ``chaining calculi'' for handling  orders 
have been defined by Bachmair and Ganzinger in \cite{BGcade,BGjacm}. 
However,
$\MC$ turns out to be quite appropriate for our context. (In
particular, since the problem of ``variable chaining'' does
not occur for the sets of clauses considered here).
\end{rem}

\begin{exa} (Example \ref{ex:delta} continued) 
According to Theorem~\ref{th:MC} we should add the sets of clauses
$\EEf$, $\EP$, and $\DO$ to $\CLval{\det p(f(x)) \Impl p(x)}$,
in order to guarantee that Irreflexivity Resolution and Chaining 
suffice to derive the empty clause,
% using only Irreflexivity Resolution and Chaining, 
witnessing the validity of $\E x (\det p(f(x) \Impl p(x))$ and consequently 
also of $\E x\A y (\det p(y) \Impl p(x))$. 
However it turns out that only the following subset of clauses is actually
needed for this purpose:
\begin{desCription}
\item\noindent{\hskip-12 pt\bf From $\CLval{\det p(f(x)) \Impl
    p(x)}$:}\ ~\\
    $\phantom01: \{p_1(x) \le \bot, \top \le p(f(x))\}$\\
    $\phantom04: \{\top\le p_2(x), p(x)<p_1(x) \}$\\ 
    $\phantom07: \{p_2(x) < \top \}$ 
\item\noindent{\hskip-12 pt\bf From $\EP$:}\ ~\\
    $E_1: \{\bot \le y\}$\\
    $E_2: \{y\le \top\}$
\end{desCription}
The empty clause can be derived as follows:\\[1ex]
\hspace*{3ex} $8: \{\top\le \top, p(x)<p_1(x) \}$ from chaining $7$ and $4$\\
\hspace*{3ex} $9: \{p(x)<p_1(x) \}$ by irreflexivity resolving $8$\\
\hspace*{2ex} $10: \{p_1(x) < \bot, \top \le p(f(x))\}$ from chaining $1$ and $9$\\
\hspace*{2ex} $11: \{\bot<\bot, \top\le p(f(x))\}$ from chaining $10$ and $E_1$\\
\hspace*{2ex} $12: \{ \top \le p(f(x))\}$ by irreflexivity resolving $11$\\
\hspace*{2ex} $13: \{\top < p_1(f(x))\}$ from chaining $12$ and $9$\\
\hspace*{2ex} $14: \{\top \le \top\}$ from chaining $13$ and $E_2$\\
\hspace*{2ex} $15: \{\}$ by irreflexivity resolving $14$
\end{exa}

%%%%%%%%%%%%%%%%%%%%%%
\section{Conclusion}

We took up the challenge of providing logical foundations for efficient theorem
proving for $\Gd$, i.e., G\"odel logic augmented by the projection
operator~$\det$. In contrast to classical logic, testing validity of a $\Gd$-formula~$F$
is not equivalent to testing the (1-)unsatisfiability of $\neg F$. However
both problems are important in view of intended applications. Unfortunately,
efficient proof search methods for unrestricted first-order~$\Gd$ seem, at least currently,
to be out of reach. In particular, Skolemization for full~$\Gd$, even without~$\det$,
is an open problem.
Consequently, we have focused on the (still very expressive)
prenex fragment of $\Gd$ and described a proof search method
that remains as close as possible in spirit to resolution 
based theorem proving for classical logic. In particular
this allows us to treat both problems, testing validity and testing 1-unsatisfiability,
in a uniform manner. While, as we have shown by proving a version of
Herbrand's theorem, standard
Skolemization preserves validity for prenex~$\Gd$, we had to come up with a novel, extended form of Skolemization for satisfiability.
In both cases, Skolemized formulas are efficiently translated into
a specific structural normal form. This consists of
sets of order clauses, where the literals are of the from $s<t$ or $s\le t$. 
We have finally explained how chaining resolution, 
a well investigated proof search method for order clauses, 
can be employed to check unsatisfiability in a machine oriented manner.  

We like to emphasize that our results not only provide a basis for automated
proof search, but also demonstrate a number of interesting logical
properties of~$\Gd$ that distinguish it, e.g., from other fuzzy logics~\cite{hajek}.
For example, the fact that the set of 1-unsatisfiable prenex formulas
of~$\Gd$ is recursively enumerable (as trivially implied by our results)
has not been known previously.

\section*{Acknowledgement}

We thank the referees of this paper and Daniel Weller for valuable remarks that
lead to considerable improvements.

%%%%%%%%%%%%%%%%%%%%%%


\begin{thebibliography}{10}

\bibitem{avr91}
A. Avron.
Hypersequents, logical consequence and intermediate logics for
concurrency. {\em Annals of Mathematics and Artificial Intelligence} 4: 
225--248, 1991.

\bibitem{baaz96} M. Baaz.
Infinite-valued G\"odel logics with
0-1-projections and relativizations. In {\em Proceedings G\"odel 96. Kurt
G\"odel's Legacy}. Springer LNL~6, 23--33,  1996.

\bibitem{BCF01}
M. Baaz, A. Ciabattoni and C.G. Ferm{\"u}ller.
Herbrand's Theorem for prenex {G}{\"o}del logic
and its consequences for theorem proving.
{\em Proceedings of LPAR'2001}. Springer LNAI 2250, 201--216, 2001.

\bibitem{BCF03} M. Baaz, A.  Ciabattoni and C.G.  Ferm\"uller.
Hypersequent calculi for G\"odel Logics --- a survey. 
{\em  J. of Logic and Computation} 13, 1-27, 2003.

\bibitem{BCP09}
M. Baaz, A. Ciabattoni and N. Preining. SAT in monadic G\"odel logics:
a borderline between decidability and undecidability. {\em Proceedings of
WOLLIC 2009}. Springer LNAI 5514, 113-123, 2009.

\bibitem{BHW}
M. Baaz, S. Hetzl and D. Weller. On the complexity of proof deskolemization. {\em J.~Symbolic Logic}. To appear. 

\bibitem{Handbook}
M. Baaz, U. Egly and A. Leitsch.
Normal form transformations. {\em Handbook of Automated Reasoning}, Vol.~1.
Eds: A.~Robinson, A.~Voronkov. Elsevier, 2001,  273--333.


\bibitem{BF10}
M. Baaz and C.G. Ferm{\"u}ller. A resolution mechanism for prenex G\"odel logic.
In {\em Proceedings of CSL 2010}. Springer LNCS 6247, 67-79, 2010. 


\bibitem{HBMFL_goedel}
M. Baaz and N. Preining. {G}\"odel-Dummett logics. 
{\em Handbook of Mathematical Fuzzy Logic}, volume~2, Eds: P.~Cintula,
P.~H\'ajek, C.~Noguera. College Publications, pp. 585--627, 2011.

\bibitem{BPZ07}
M. Baaz, N. Preining and R. Zach.
First-order {G}\"odel logics.
{\em Annals of Pure and Applied Logic} 147:23--47, 2007.

\bibitem{BGcade} L. Bachmair and H. Ganzinger.
Ordered chaining for total orderings. {\em Proc. CADE`94}, Springer LNCS 814,
1994,  435--450.

\bibitem{BGjacm} L. Bachmair and H. Ganzinger.
Ordered chaining calculi for first-order theories of transitive relations.
{\em J.\ ACM}  45(6): 1007-1049, 1998.

\bibitem{CR10}
A. Ciabattoni and P. Rusnok. On the classical content of monadic G$^\sim$ and
its application to a fuzzy medical expert system, {\em Proceedings of KR 2010},
AAAI, 373--381, 2010.

\bibitem{DV} A. Degtyarev and A. Voronkov. Decidability problems for the prenex
fragment of intuitionistic logic. In {\em Proceedings LICS'96}. IEEE Press,
503-509, 1996.

\bibitem{dummett} M. Dummett. A propositional calculus
with denumerable matrix. {\em J.~Symbolic Logic} 24: 97--106, 1959.

\bibitem{Gab72} D.M. Gabbay. Decidability of some intuitionistic
predicate theories.
{\em J.~of Symbolic Logic} 37: 579--587, 1972.

\bibitem{Fiorino} G. Fiorino. Fast decision procedure for 
propositional Dummett logic based on a multiple premise tableau 
calculus. {\em Inf. Sci.}~180(19): 3633--3646, 2010.

\bibitem{hajek} P. H\'{a}jek. {\em Metamathematics of Fuzzy Logic}.
Kluwer, 1998.

\bibitem{hajek1} P. H\'{a}jek. Arithmetical complexity of fuzzy predicate logics
 -- a survey. {\em Soft Computing}~9/12: 935--941, 2005.

\bibitem{hajek2} P. H\'{a}jek. Arithmetical complexity of fuzzy predicate logics
 -- a survey~II. {\em Annals of Pure and Applied Logic}~161(2): 212--219, 2009.

\bibitem{HB} D. Hilbert and P. Bernays. {\em Grundlagen der Mathematik} , Vol. 2, Berlin: Springer
1939.

\bibitem{LW07} D. Larchey-Wendling. Graph-based decision for 
G\"odel-Dummett logics. {\em J. Autom. Reasoning}~38(1-3): 201--225, 2007.


\bibitem{leitsch} A. Leitsch. {\em The Resolution Calculus}. 
Springer (fomerly Kluwer) 1997.

\bibitem{LPV}
V. Lifschitz, D. Pearce and A. Valverde. Strongly equivalent logic programs.
{\sl ACM Transaction on Computational Logic}, 2(4): 526--541, 2001.

\bibitem{Mint} G. Mints. The Skolem method in intuitionistic calculi. 
{\em Proceedings
of Steklov Institute of Mathematics}, 121: 73--109, 1972.

\bibitem{PG} D. Plaisted and  S. Greenbaum.
A structure-preserving clause form translation.
{\em J.~Symbolic Computation} 2: 293--304, 1986.


\bibitem{P10}
N. Preining. G\"odel logics - a survey. {\em Proceedings of LPAR 2010}. Springer LNCS 6397, pp. 30--51, 2010.


\bibitem{TT84} G. Takeuti and T. Titani. Intuitionistic fuzzy logic
and intuitionistic fuzzy set theory. {\em J.~of Symbolic Logic},
49: 851--866, 1984.

\bibitem{visser}
A. Visser. On the completeness principle: a study of provability in
Heyting's Arithmetic. {\sl Annals of Math. Logic} 22: 263--295, 1982.

\end{thebibliography}
\end{document}